\documentclass[aps,prb,floatfix,twocolumn,superscriptaddress]{revtex4}

\usepackage{color,graphicx}

\newcommand{\mub}{$\mu_{\rm B}$}
\newcommand{\TN}{T$_{\rm N}$}

\begin{document}

\title{Long period helical structures and twist-grain boundary phases induced by non magnetic ion doping in Mn$_{1-x}$(Co,Rh)$_{x}$Ge chiral magnet}

\author{N. Martin}\email[]{nicolas.martin@cea.fr}\affiliation{Laboratoire L\'eon Brillouin, CEA, CNRS, Universit\'e Paris-Saclay, CEA Saclay 91191 Gif-sur-Yvette, France}
\author{M. Deutsch}\affiliation{Universit\'e de Lorraine, Laboratoire CRM2,UMR UL-CNRS 7036, 54506 Vandoeuvre-les-Nancy,France}
\author{G. Chaboussant}\affiliation{Laboratoire L\'eon Brillouin, CEA, CNRS, Universit\'e Paris-Saclay, CEA Saclay 91191 Gif-sur-Yvette, France}
\author{F. Damay}\affiliation{Laboratoire L\'eon Brillouin, CEA, CNRS, Universit\'e Paris-Saclay, CEA Saclay 91191 Gif-sur-Yvette, France}
\author{P. Bonville}
\affiliation{SPEC, CEA, CNRS, Universit\'e Paris-Saclay, CEA-Saclay, 91191 Gif-sur-Yvette, France}
\author{L. N. Fomicheva}
\affiliation{Vereshchagin  Institute for  High Pressure Physics, Russian Academy of Sciences, 142190, Troitsk, Moscow, Russia}
\author{A. V. Tsvyashchenko}\affiliation{Vereshchagin  Institute for  High Pressure Physics, Russian Academy of Sciences, 142190, Troitsk, Moscow, Russia}
\affiliation{Skobeltsyn Institute of Nuclear Physics, MSU, Vorob'evy Gory 1/2, 119991 Moscow, Russia}
\author{U.K. R\"ossler}\affiliation{IFW Dresden, PO Box 270116, 01171 Dresden, Germany}
\author{I. Mirebeau}\affiliation{Laboratoire L\'eon Brillouin, CEA, CNRS, Universit\'e Paris-Saclay, CEA Saclay 91191 Gif-sur-Yvette, France}

\date{\today}

\begin{abstract}
We study the evolution of helical magnetism in MnGe chiral magnet upon partial substitution of Mn for non magnetic 3d-Co and 4d-Rh ions. At high doping levels, we observe spin helices with very long periods -more than ten times larger than in the pure compound- and sizable ordered moments. This behavior calls for a change in the energy balance of interactions leading to the stabilization of the observed magnetic structures. Strikingly, neutron scattering unambiguously shows a double periodicity in the observed spectra at $x \gtrsim 0.45$ and $\gtrsim 0.25$ for Co- and Rh-doping, respectively. In analogy with observations made in cholesteric liquid crystals, we suggest that it reveals the presence of magnetic twist-grain-boundary phases, involving a dense short-range correlated network of screw dislocations. The dislocation cores are described as smooth textures made of non-radial double-core skyrmions.  
\end{abstract}


\maketitle


In condensed matter, effects of quenched disorder\cite{anderson1979ill} on an ordered phase can be strikingly different, varying 
between marginal modifications of phase transitions to the destruction of homogeneous order and the occurrence of disordered, amorphous or glassy states\cite{young1997spin}. 
Particularly interesting are systems where disorder only partly destroys the ordered state and prompts the appearance of localized defects. Liquid crystals and helium quantum liquids confined in random environments are well studied examples of such systems\cite{bellini2001universality,volovik2003universe}.

Chiral helimagnets are presently at the focus of much interest. In pure form,  they are incommensurate spiral magnets but can also form double-twisted solitonic states -now known as {\it skyrmions}- on length scales much larger than lattice spacings. These modulated spin textures are akin to cholesteric liquid crystals. The helimagnetic state in the long-range limit is also closely related to one-dimensional smectic liquid crystals\cite{radzihovsky2011nonlinear}. The free energy of this state is usually described in a continuum model by a functional involving an effective exchange constant $\mathcal{A}$ and two anisotropy terms with Dzyaloshinskii-Moriya (DM) interaction $\mathcal{D}$ and exchange anisotropy $\mathcal{B}$, respectively. A hierarchical model\cite{Bak1980} leads to a helical magnetic ground state with period $\lambda \simeq \mathcal{A}/\mathcal{D}$. In a random system, $\mathcal{A}$, $\mathcal{D}$ and $\mathcal{B}$ are spatially varying but with fixed distributions. Thanks to proper metric rescaling of space, quenched disorder can formally be canceled, as long as it does not lead to an inversion of the locally favored twisting. Its leading effect comes from random anisotropy, so that a ground state helical structure should basically be preserved, while randomness would only modify the local wavelength, rotation axis and, eventually, propagation direction of spin spirals. However, the equivalence of the long-distance behavior of the helimagnetic ground state with smectic liquid crystals allows us to use basic theories about smectics in random anisotropic media\cite{radzihovsky1997dirt,radzihovsky1997nematic,radzihovsky1999smectic,bellini2001universality}, which assert that any anisotropic disorder destroys long range order. It suggests that, as in smectic crystals, spin helices in chiral systems could be modified by the penetration of dislocations\cite{radzihovsky1999smectic}, the density of which increases with the amount of disorder, yielding twist-grain-boundary (TGB) phases\cite{renn1988abrikosov}. 

Here, we use alloys of the cubic magnet MnGe to investigate the influence of quenched disorder on the chiral helimagnetic ground-state. 
We have substituted Mn for 3d-Co or 4d-Rh non magnetic ions
and focused on compositions belonging to the Mn rich side (x $\leq$ 0.5) of the Mn$_{1-x}$(Co,Rh)$_{x}$Ge series. In both cases, we observe that doping 
induces helical structures with very long periods (up to $\simeq$ 500 \AA, as compared to $\simeq 30$ \AA~for pure MnGe\cite{Makarova2012}) and sizable ordered moments. Moreover, 
we show that they differ from harmonic helices by the presence of $\it{two}$ magnetic diffraction peaks, calling for an additional periodicity. We propose that it reveals the presence of magnetic TGB phases, as detailed in the last part of this letter. 

\begin{center}
\begin{figure}[!ht]
\includegraphics[width=8.5cm]{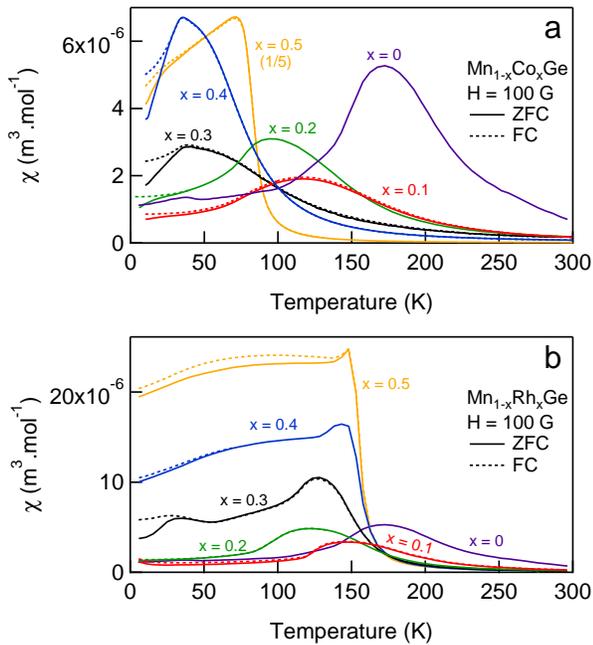}
\caption{\label{fig:dc_susceptibility} Temperature dependence of the static magnetic susceptibility of  Mn$_{1-x}$Co$_{x}$Ge \textbf{(a)} and Mn$_{1-x}$Rh$_{x}$Ge \textbf{(b)} obtained in zero field cooled (ZFC) and field cooled (FC) configurations at an applied field of 100 G. In panel (a), $x = 0.5$ curve is divided by a factor 5 to ease visual inspection.}
\end{figure}
\end{center}

%
For all studied samples, the low field static susceptibility $\chi$ follows a Curie-Weiss law at high temperature\cite{SuppMat}. 
In both compounds, for $x \leq 0.2$, one observes a broad antiferromagnetic-like (AFM) peak, akin to that occuring in MnGe around \TN\ $= 170$ K\cite{Kanazawa2011,Deutsch2014b,Altynbaev2014,Viennois2015}, which moves towards lower temperatures with increasing Co-doping (Fig. \ref{fig:dc_susceptibility}a) while its position remains fairly constant in the Rh-doped system (Fig. \ref{fig:dc_susceptibility}b).
For $x \geq 0.3$, $\chi$ shows a ferromagnetic-like (FM) upturn and its magnitude strongly increases with increasing $x$. At lower temperatures, irreversibilities appear between field-cooled (FC) and zero-field-cooled (ZFC) curves. As will be discussed later, the transition from AFM to FM regime, as $x$ increases beyond $\simeq 0.3$, is associated with the stabilization of long period (LP) helimagnetic structures.

\begin{center}
\begin{figure}[t]
\includegraphics[width=8.5cm]{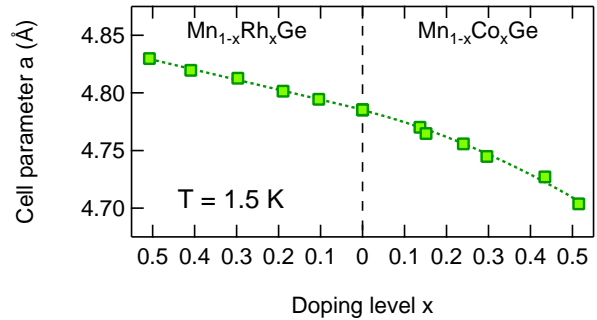}
\caption{\label{fig:lattice_params}Low temperature cubic lattice constant of Mn$_{1-x}$(Co,Rh)$_x$Ge versus doping $x$ determined by neutron powder diffraction.}
\end{figure}
\end{center}

\begin{figure*}[!ht]
\center{\includegraphics[width=\textwidth]{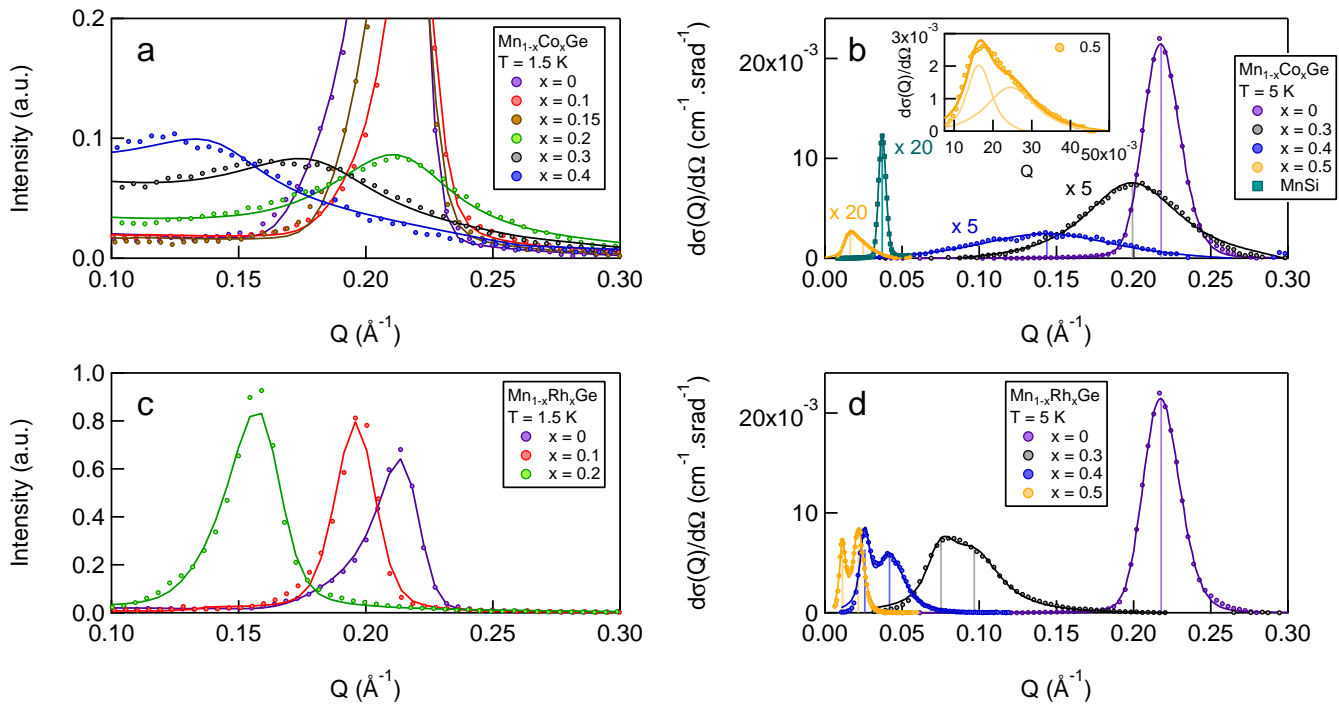}}
\caption{\label{fig:MnCoRhGe_diff_and_SANS} Evolution of the helimagnetic peak in Mn$_{\rm 1-x}$Co$_{\rm x}$Ge \textbf{(a)} and Mn$_{\rm 1-x}$Rh$_{\rm x}$Ge \textbf{(c)} upon doping at T $= 1.5$ K, as measured by neutron powder diffraction. For each concentration a pattern measured at 300\,K was subtracted to eliminate background. For display purposes, patterns were calibrated to the intensity of the (110) nuclear reflection. Small-angle neutron scattering patterns of Mn$_{\rm 1-x}$Co$_{\rm x}$Ge \textbf{(b)} and Mn$_{\rm 1-x}$Rh$_{\rm x}$Ge \textbf{(d)} taken at T = 5 K. On can note \emph{(i)} the large increase of the helical wavelength as a function of $x$ and \emph{(ii)} the emergence of a second peak for $x \geq 0.5$ (Co) (see also inset of \textbf{(b)}) and $x \geq 0.3$ (Rh). In all panels, solid lines are fit curves as described in the text (namely, outcome of a Rietveld refinement of the powder diffraction patterns in \textbf{(a)} and \textbf{(c)} and fit of a convolution product of a Lorentzian scattering function with the SANS Gaussian resolution function in \textbf{(b)} and \textbf{(d)}).}
\end{figure*}

In order to follow the evolution of the magnetic structure in Mn$_{1-x}$(Co,Rh)$_{x}$Ge, neutron powder diffraction measurements were performed on the two-axis instrument G4.1 at the Laboratoire L\'eon Brillouin (LLB) using a neutron wavelength $\lambda = 2.428$ \AA. Interestingly, Co and Rh substitution respectively contracts and expands the cubic MnGe lattice (Fig. \ref{fig:lattice_params}). The obtained powder patterns show satellites of the nuclear Bragg reflections as a hallmark of helimagnetic long-range ordering. In pure MnGe, the intense satellite of the $Q = 0$ Bragg peak, clearly visible at low angles, coexists with much weaker satellites at larger angles\cite{Makarova2012}. Its evolution with increasing Co content is shown in Fig. \ref{fig:MnCoRhGe_diff_and_SANS}a at base temperature. With increasing $x$, the helimagnetic satellite broadens, its intensity strongly decreases and its position moves towards lower Q values. For $x = 0.5$, the peak has evaded from the diffraction window. In Mn$_{\rm 1-x}$Rh$_{\rm x}$Ge, the intensity of the satellite decreases less rapidly, but is position shifts very quickly with increasing $x$ and it disappears from the diffraction window for $x$ \textgreater \, 0.2 (Fig. \ref{fig:MnCoRhGe_diff_and_SANS}c).

In both cases, measurements covering a lower Q range are thus needed to follow the evolution of magnetic structure. In that respect, small-angle neutron scattering (SANS) is an ideal method since it can access momentum transfers as small as few 10$^{-4}$ \AA $^{-1}$. Experiments were performed on the SANS spectrometers PA20 and PAXY\cite{Chaboussant2012a,Chaboussant2012b} of the LLB, using incident wavelengths 4.46 and 6 \AA, and sample-to-detector distances ranging between 2 and 10 m. Spectra were corrected for detector efficiency and calibrated cross sections were obtained by taking sample thickness and transmission, as well as incident neutron flux, into account\cite{SuppMat}. In Mn$_{\rm 1-x}$Co$_{\rm x}$Ge, the helimagnetic peak is now clearly evidenced for $x = 0.5$ (Fig. \ref{fig:MnCoRhGe_diff_and_SANS}b). Its asymmetric lineshape is best accounted for by a sum of two peaks (inset of Fig. \ref{fig:MnCoRhGe_diff_and_SANS}b). In Mn$_{\rm 1-x}$Rh$_{\rm x}$Ge, we observe the helimagnetic peak for $x$ \textgreater\, 0.2 (Fig. \ref{fig:MnCoRhGe_diff_and_SANS}d). Moreover, thanks to the symmetric and narrow line shape of the resolution function, we observe a resolved two peak structure. The appearance of a second peak calls for a new periodicity in the system occurring at high doping, namely at $x$ \textgreater\,$0.2$ and $x = 0.5$ for Rh and Co substitutions respectively\cite{NoteDoublePeak}.

\begin{center}
\begin{figure}[t]
\includegraphics[width=8.5cm]{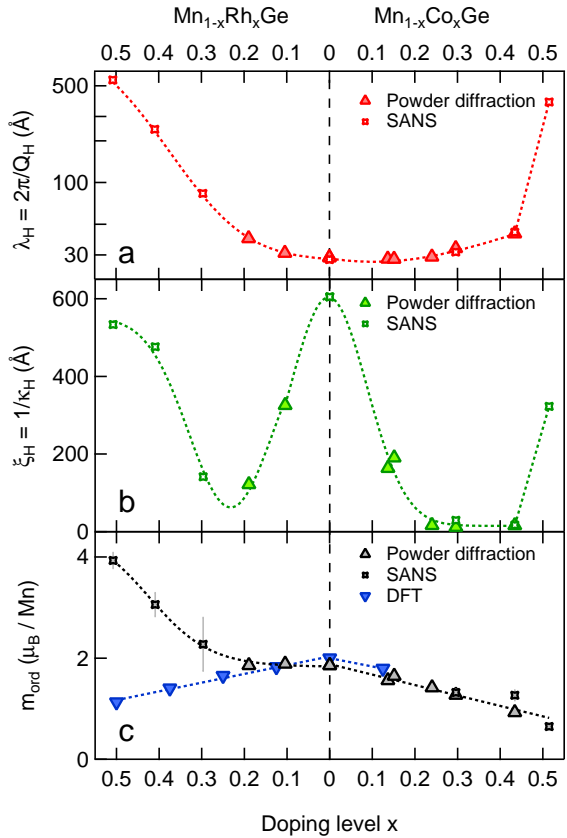}
\caption{\label{fig:MnCoRhGe_pitch} Doping dependence of the helical wavelength \textbf{(a)}, coherence length \textbf{(b)} and ordered magnetic moment \textbf{(c)} in Mn$_{\rm 1-x}$(Co,Rh)$_{\rm x}$Ge as determined by powder neutron diffraction and small-angle scattering (see text). In panel \textbf{(c)}, results of \emph{ab initio} calculation of the local moment performed on a 2x1x1 supercell are shown for comparison.}
\end{figure}
\end{center}

In the data treatment, we assumed helical order with magnetic moments on the Mn sites only, propagating along (001) axes, as for pure MnGe. Diffraction patterns were described in the cubic space group $P2_{1}3$, with a helical wavevector $\mathbf{Q_{\rm H}} = (0,0,Q_{\rm H})$, using the Fullprof suite\cite{Fullprof1993}. Combined nuclear and magnetic refinements provides the helical wavenumber $Q_{\rm H}$, inverse correlation length $\kappa_{\rm H}$ and value of the Mn ordered moment $m_{\rm ord}$. 
In the SANS case, parameters describing magnetic order are obtained by assuming a Ornstein-Zernike form for the scattering functions $\mathcal{S}(Q)$ (with peak position $Q_{\rm H}$ and half-width at half-maximum $\kappa_{\rm H}$), convolved with the calculated resolution function $\mathcal{R}(Q)$\cite{SuppMat}. We have calibrated $m_{\rm ord}$ on an absolute scale by measuring the integrated intensity of the helical peaks of MnGe ($m_{\rm ord}$ = 1.85 \mub / Mn) and MnSi ($m_{\rm ord}$ = 0.4 \mub / Mn) powder samples\cite{SuppMat}. This analysis strategy is devoid of any \emph{a priori} hypothesis on the nature of the observed long-range order and is found to reproduce very well magnetic SANS patterns for all studied compositions. The helix pitch $\lambda_{\rm H} = 2\pi/Q_{\rm H}$, coherence length $\xi_{\rm H} = 1/\kappa_{\rm H}$ and Mn moment $m_{\rm ord}$ observed at base temperature are reported in Fig. \ref{fig:MnCoRhGe_pitch}. 

The first striking feature is the huge increase of the helix pitch above $x \simeq 0.3$, which reaches 380 and 550 \AA \, at $x =0.5$ in Co and Rh systems, respectively. The huge increase of the helical period calls for of a deep change in the balance of magnetic interactions. It is indeed likely that frustrated next nearest neighbour AFM interactions could play a major role in stabilizing the short helical pitch in pure MnGe\cite{Chizhikov2013}. As a primary effect, doping with non magnetic ions suppresses these interactions by breaking mostly the numerous next nearest neighbor Mn-Mn bonds. At high doping, this results in an increase of the average FM exchange with respect to the DM anisotropy, so that the hierarchical model\cite{Bak1980} at play in weak itinerant ferromagnets such as MnSi or FeGe starts to be valid.

\begin{center}
\begin{figure}[t]
\includegraphics[width=8.5cm]{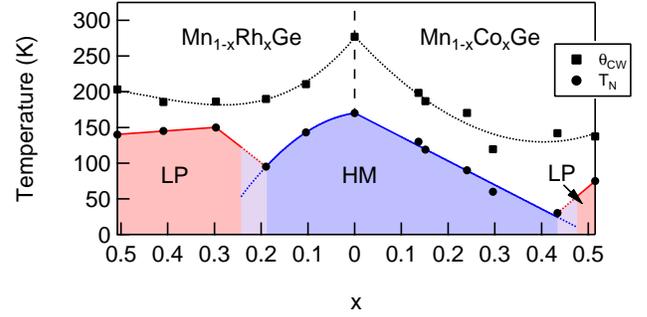}
\caption{\label{fig:PhaseDiagram} Magnetic phase diagram of Mn$_{1-x}$(Co,Rh)$_{x}$Ge inferred from static susceptibility (Curie-Weiss temperature $\theta_{\rm CW}$) and neutron scattering (N\'eel temperature, T$_{\rm N}$). At low temperature, below $x \simeq 0.45$ for Co doping and $x \simeq 0.25$ for Rh doping, a simple helimagnetic (HM) state is observed. At higher values of $x$, supermodulations of the harmonic helix are observed (LP).}
\end{figure}
\end{center}

Additionaly, it is instructive to note that, after an initial decrease, the coherence length shows an upturn at concentrations where long periods and double peaks appear, suggesting that the LP structures are ordered over very large length scales. 
Finally, the $x$-dependence of the ordered moment and of the N\'eel temperatures \TN \, deserves attention. In Mn$_{1-x}$Co$_{x}$Ge, both quantities are found to decrease with increasing x, as expected from the substitution of Mn by non magnetic ions (Fig. \ref{fig:PhaseDiagram}). The situation is notably different in Rh-doped MnGe which, instead, shows a stable N\'eel temperature and an increasing ordered moment as a function of $x$. {\it Ab initio} calculations of the local moment in both systems underscore the abnormality of such a behaviour. Indeed, the experimental points clearly deviate from the expected linear decay of the moment as a function of $x$, actually found by DFT in the case of Co-doping. Chemical pressure therefore  plays an important role here, since lattice expansion stabilizes higher moments and transition temperatures in Rh than in Co systems for the same amount of doping. This is in fact in agreement with recent observations of pressure-induced spin transitions in pure MnGe\cite{Deutsch2014a,Martin2016b} which established the presence of a strong magneto-elastic coupling in this system.

The magnetic phase diagram of Mn$_{1-x}$(Co,Rh)$_{x}$Ge ($x \leq 0.5$) is shown in Fig. \ref{fig:PhaseDiagram}. It demonstrates that Co-doping is more prone to destabilizing long-range order than its Rh counterpart, as seen by the compared $x$-dependence of ordering $T_{\rm N}$ and Curie-Weiss $\theta_{\rm CW}$ temperatures, respectively inferred form neutron scattering and the high temperature linear slope of $\chi^{-1}$\cite{SuppMat}. In both cases, however, a transition is observed from a helimagnetic state towards a new type of magnetic structure, as evidenced by an additional peak appearing in the SANS patterns. 

We now describe this double periodicity shown by the LP structures at high doping\cite{NoteCoRhrichside}. It is observed in all cases, suggesting an intrinsic character. The second peak is not a high order satellite of the helical order, excluding a description by anisotropy induced square modulations of the local magnetization\cite{Izyumov1984}, as well as multiple Bragg scattering. Our refinements of the crystal structure also exclude a macroscopic phase separation, since all samples remain in a single cubic phase without noticeable broadening of the measured nuclear peaks. 
Noting that the two peaks of the LP structures vary with T and x in correlated ways, we propose that quenched disorder occurs in the shape of spatially correlated regions. This long-period spin arrangement is now addressed under a theoretical viewpoint.

\begin{figure*}[!ht]
\center{\includegraphics[width=0.8\textwidth]{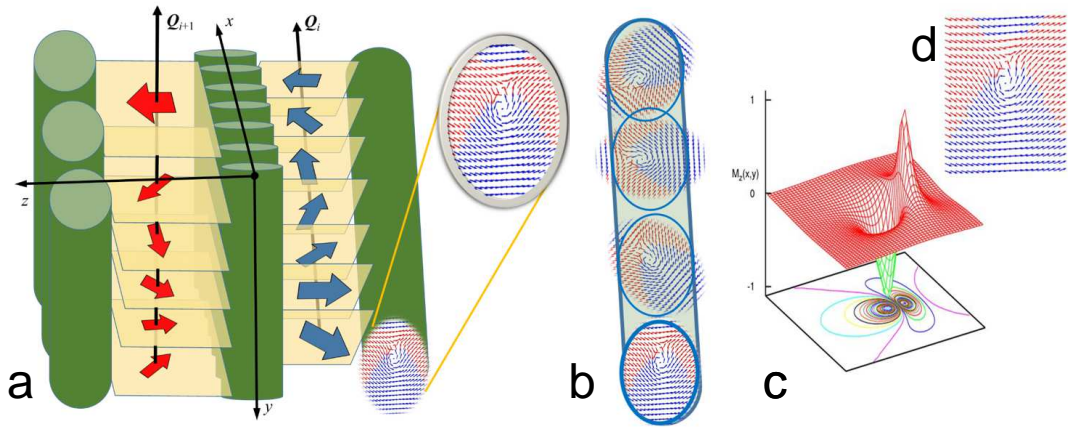}}
\caption{\label{fig:TGB}\textbf{(a)} Schematic view of the proposed twist grain boundary (TGB) structure in a chiral helimagnet. \textbf{(b)} Internal structure of a screw dislocation made of stacked double core skyrmion. \textbf{(c)} and \textbf{(d)} Cross section of a double core skyrmion: \textbf{(c)} out of plane magnetization component M$_{\rm Z}$ and \textbf{(d)} projection of the in-plane component, showing a narrowly coupled vortex-antivortex configuration.
}
\end{figure*}

As observed in smectic liquid crystalline systems, quasi-crystalline dislocations can occur in directions perpendicular to the propagation direction of the primary 1D modulation in chiral systems realizing TGB phases\cite{renn1988abrikosov}. Along the propagation direction of the primary spiral order the balance between exchange and twisting DM-term is optimal, but the spiral order is still frustrated and, upon insertion of dislocations, may twist additionally in the perpendicular direction (see Fig. \ref{fig:TGB}a for a schematic drawing of this phase). These textures are caused by the defect generation through 'penetration of chirality' described by de Gennes in analogy to type II superconductors\cite{deGennes1972analogy}. A planar arrangement of parallel screw-dislocations  (in $xy$-plane, Fig. \ref{fig:TGB}) connects slab-like regions of spiral helices with differing propagation direction, thus, it acts as a TGB. Periodic arrangements of these TGBs can form a supermodulation of the primary 1D spiral texture, which is seen by the occurrence of additional Bragg peaks in a diffraction experiment.

However, the proposed structure should differ in typical lateral lengths from the TGB-phases typically observed in chiral smectics, owing to the differing structure of order parameter and couplings. The lateral twisting of the DM spiral and the periodic insertion of screw dislocations should take place on the length scale of the twisting period $\lambda_{\rm H}$ itself. Therefore, the dislocation cores should occupy a large fraction of the volume, as supported by the relatively large intensity of the secondary peak observed by SANS. The internal structure should be smooth and defect-free in terms of the  ferromagnetic order, which means that the high energy cost of forming singular hedgehog-defects (Bloch points) is clearly to be avoided. Therefore, in the core of the screw dislocations an escape of the ferromagnetic vector $\mathbf{m}$ in the third dimension and a full sweep over the $4\pi$ surface of the sphere as order-parameter manifold takes place.

The dislocation core is a smooth texture described by a non-radial double-core skyrmion (Fig. \ref{fig:TGB}). The projection of $\mathbf{m}$ forms a vortex-antivortex-pair in the plane of the cross-section perpendicularly to the dislocation line. At the locations of the vortex-singularities $\mathbf{m}$ is exactly perpendicular to the plane pointing up and down. Along the dislocation line, the locations of these two extremal values of $\mathbf{m}$ twist in a cork-screw fashion around each other. Analogous dislocation structures have been observed and studied in cholesteric liquid crystals \cite{bouligand1970paires,rault1971lignes,rault1973dislocation,rault1974dislocation}. Recent simulation studies on slabs of cubic helimagnets and on smooth-line-like textures of conical helix states report such 3D-twistings. They correspond to a partial penetration of chirality into the 1D spiral, related either to the surface cut \cite{rybakov2013three}, or to non-radial skyrmion excitation of the homogeneous smectic-like states, which requires a modulation along the tubular line-like texture \cite{leonov2016three}.

In summary, non magnetic ion substitution in B20-alloys Mn$_{1-x}$(Co,Rh)$_x$Ge induces magnetic structures with very long periods and magnetic moments strongly dependent on the unit cell volume. While the first feature reveals a change in the energy balance of magnetic interactions, from AFM frustrated nearest neighbour to the hierarchical Bak-Jensen scheme, the second one suggests deep modifications of the material band structure.  The anomalous lineshape observed in SANS at high doping supports the general assumption that strong quenched disorder is able to partly destroy the spiral ground-state and stabilize a 3D TGB-like phase with a very dense and short-range correlated network of screw dislocations. Each dislocation is a double-core singly charged skyrmion. The whole structure is thus topologically non trivial, although the net topological charge is zero. Transitions from simple spin spirals to TGB structures occur at $x \simeq 0.45$ for Mn$_{1-x}$Co$_{x}$Ge and $x \simeq 0.25$ for Mn$_{1-x}$Rh$_{x}$Ge. Whether this process takes place through a quantum critical point or in a thermally assisted way remains to be checked experimentally.

We thank E. Altynbaev and S. Grigoriev for very insightful discussions, S. Gautrot and V. Thevenot for technical assistance and A. Bauer for provision of MnSi powder samples. The post doc training of N. Martin was funded by the LabEx Palm. Public grant from the "Laboratoire d'Excellence Physics Atom Light Mater" (LabEx PALM) overseen by the French National Research Agency (ANR) as part of the "Investissements d'Avenir" program (ANR-10-LABX-0039). The post doc training of M. Deutsch was funded by the ANR (DYMAGE). L.N. Fomicheva and A.V. Tsvyashchenko acknowledge support of the Russian Foundation for Basic Research (Grant 17-02-00064). 



\begin{thebibliography}{32}
\bibitem[{Not({\natexlab{a}})}]{NoteDoublePeak}
\bibinfo{howpublished}{Note that when two magnetic diffraction peaks are
  observed, we choose the one located at the lowest $Q$ to represent long-range
  helical order. This choice is motivated later, assuming that screw
  dislocations are embedded in a helimagnetic background and therefore must
  possess a smaller spatial periodicity.}
\bibitem[{Not({\natexlab{b}})}]{NoteCoRhrichside}
\bibinfo{note}{Preliminary measurements show that such
  structures, with extremely long periods and sizable moments, are also observed in
  both systems on the Co- and Rh-rich side.}
\end{thebibliography}
\end{document}